\documentclass[prl,aps,showpacs,preprint]{revtex4}
\def\bvec#1{{\rm\bf #1}}

\def\bea{\begin{eqnarray}}
\def\eea{\end{eqnarray}}
\def\be{\begin{equation}}
\def\ee{\end{equation}}

\begin{document}
\title{Simple solutions of fireball hydrodynamics
for self-similar, ellipsoidal flows}
\author{T. Cs{\"o}rg{\H o}}
\affiliation{Dept. Phys. Columbia University, 
538W 120th St, New York, NY - 10027, USA, \\
MTA KFKI RMKI, H-1525 Budapest 114, POB 49, Hungary} 
\email{csorgo@sunserv.kfki.hu}
\pacs{24.10.Nz,47.15.Hg}
\begin{abstract}
A new family of simple, analytic solutions of self-similarly 
expanding fireballs is found for systems with ellipsoidal symmetry and a direction dependent,
generalized Hubble flow.  Gaussian, shell like or oscillating density profiles emerge 
for simple choices of an arbitrary scaling function.  
New, cylindrically or spherically symmetric as well as approximately
one dimensional hydrodynamical solutions are obtained for various special
choices of the initial conditions.
\end{abstract}
\maketitle

{\it Introduction} --- 
Hydrodynamics is describing the local conservation of matter,
momentum and energy. Due to this nature, hydrodynamical solutions
are applied to a tremendous range of physical phenomena ranging from the
stellar dynamics to the description of high energy
collisions of heavy ions  as
well as collisions of elementary particles. Some of the 
most famous hydrodynamical solutions, like the Hubble flow of our 
Universe or the Bjorken flow in ultra-relativistic heavy ion
physics  have the properties of self-similarity  and 
scale-invariance. Heavy ion collisions are known to create three
dimensionally expanding systems. In case of non-central collisions,
cylindrical symmetry is violated, but an {\it ellipsoidal} symmetry can
be well assumed to characterize the final state.
The data motivated, spherically or cylindrically symmetric
hydrodynamical parameterizations and/or solutions 
of refs.~\cite{jnr,jde,nr,3d,qm95,nrt,cssol,cspeter,ellsol,csell}
are generalized here to the case of such an  ellipsoidal symmetry, 
providing new families of exact analytic hydrodynamical solutions.

{\it The new family of self-similar ellipsoidal solutions} --- 
The non-relativistic (NR) hydrodynamical systems are 
specified by the continuity, Euler and energy equations: 
\begin{eqnarray}
{\partial_t n} + {\rm {\bf \nabla }} (n {\rm {\bf v }}) & = & 0,
\label{e:cont} \\
{\partial_t {\rm {\bf v}}} + ({\rm {\bf v}}{\rm {\bf \nabla}}) {\rm {\bf v }}
& = & - ({\rm {\bf \nabla }} p) / (m n) ,  \label{e:Eu} \\
{\partial_t \epsilon } + {\rm {\bf \nabla }} (\epsilon {\rm {\bf v }} ) & =
& - p {\rm {\bf \nabla }} {\rm {\bf v }} .  \label{e:en}
\end{eqnarray}
Here $n$ denotes the particle number density, ${\rm {\bf v}}$ stands
for the NR flow velocity field, $\epsilon $ for the energy
density, $p$ for the pressure and in the following the temperature field is
denoted by $T$. These fields depend on the time $t$ as well as on the 
coordinates ${\rm {\bf r}} = (r_x, r_y, r_z)$.
We assume, for the sake of simplicity, the following  
equations of state, 
\begin{eqnarray}
p & = & n T,   \qquad
\epsilon \,  = \, \kappa p,  \label{e:eos}
\end{eqnarray}
which close the set of equations for $n$, ${\rm {\bf v}}$ and $T$.
The NR ideal gas corresponds to $\kappa = \frac{3}{2}$.
The new family of exact analytic solutions of 
the hydrodynamical problem are given for arbitrary values of 
$\kappa> 0$,  $m>0$. The hydro solutions are determined by the 
choice of a positive function of a non-negative, real variable 
${\cal T}(s)$, corresponding to the (dimensionless)
scaling function of the temperature.
The scaling variable $s$ is defined as 
\bea
	s & = & \frac{r_x^2}{X^2} + \frac{r_y^2}{Y^2} + \frac{r_z^2}{Z^2}.
	\label{e:scale}
\eea
There the scale parameters depend on time, $(X,Y,Z) = (X(t), Y(t), Z(t))$.
The ellipsoidal symmetry of the solutions is reflected by 
the ellipsoidal family of  surfaces  given by  the 
$s=s_0=const$  equation. The temperature and the density field depend 
on the coordinates $(r_x, r_y, r_z)$ only through the scaling variable $s$.

The new family of elliptically symmetric solutions of fireball hydrodynamics
is given by 
\begin{eqnarray}
	n(t,{\rm {\bf r}}) & = &  n_0
		{\frac{V_0} {V}} {\nu}(s),
		\label{e:n} \\
	{\rm {\bf v}}(t,{\rm {\bf r}}) &=&
    		\left(  {\frac{\dot{X}}{X}}\, r_x,
     			{\frac{\dot{Y}}{Y}}\, r_y, 
     			{\frac{\dot{Z}}{Z}}\, r_z 
		\right),  \label{e:v} \\
	T(t, {\rm {\bf r}}) & = & T_0 \left( 
	{\frac{\displaystyle\phantom{|}V_0}{\displaystyle \phantom{|}V}} 
	\right)^{1/\kappa} {\cal T}(s) ,  \label{e:T} \\
	{\nu}(s) & = &
		\frac{1}{{\cal T}(s)}
    		\mathrm{exp}\left( - \frac{T_i}{2 T_0} 
		\, \int_0^{s}  \frac{du}{{\cal T}(u)} \right),  
			\label{e:nutau}
\end{eqnarray}
where the constant $n_0$ is given by $n_0 = n(t_0, {\bf 0})$,
the  typical volume of the expanding system is $V = XYZ$, 
the initial volume being $V_0=V(t_0)$, 
the dimensionless scaling function of the
density profile is denoted by ${\nu}(s)$,
the constant $T_0$ 
is defined by $T_0=T(t_0, {\bf 0})$ and
a constant of integration is denoted by $T_i$.
 The definitions of $n_0$ and $T_0$
correspond to the normalization ${\nu}(s=0)=1$
and ${\cal T}(s=0) = 1$. 
Initially, only one of the temperature and density
profiles can be chosen as an arbitrary positive function,
the equations of state relates the density and temperature profiles,
resulting in the matching condition for the profile functions,
as expressed by eq.~(\ref{e:nutau}).

The equations of motion of the scale parameters are 
\begin{equation}
X \ddot{X} = Y \ddot{Y} = Z \ddot{Z} = \frac{T_i}{m} 
	\left( {\frac{\displaystyle
\phantom{|}V_0}{\displaystyle\phantom{|}V}}\right)^{1/\kappa} .  \label{e:lag}
\end{equation}

This time evolution of the radius parameters $X$, $Y$ and $Z$
is equivalent 
to the classical motion of a particle  in a
non-central potential, governed by the Hamiltonian  
\begin{equation}
H = \frac{1}{2m} \left( P_x^2 + P_y^2 + P_z^2\right) + 
\kappa T_i
\left( \frac{X_0 Y_0 Z_0}{X Y Z} \right)^{1/\kappa},
\end{equation}
where the canonical coordinates are $(X,Y,Z)$ and  
the canonical momenta are $(P_x, P_y, P_z) = m (\dot{X}, \dot{Y}, \dot{Z})$.
This generalizes earlier  results~\cite{ellsol,csell} from $\kappa = 3/2$
and ${\cal T}(s) \equiv 1$ to 
and to arbitrary  ${\cal T}(s)>0$ and $\kappa > 0$.

The conservation of energy 
by the classical Hamiltonian motion determines the physical meaning
of the constant of integration $T_i$ as the initial potential
energy corresponding  to the initial internal energy of the fireball:
\bea
 E_{\rm tot}  
 & = & 
	\frac{m}{2}  
		(\dot X_0^2 + \dot Y_0^2 + \dot Z_0^2) + \kappa T_i,
\eea
where the total energy is denoted by $E_{\rm tot}$, and the initial velocities
are denoted by $(\dot X_0,\dot Y_0,\dot Z_0)$. 
Due to the 
repulsive nature of the potential, the coordinates diverge for large
values of $t$. The asymptotic velocities tend to constants~\cite{ellsol}  of
$(\dot X_{\rm as},\dot Y_{\rm as},\dot Z_{\rm as})$, 
\bea
 E_{\rm tot} & = & 
	\frac{m}{2}  
		(\dot X_{\rm as}^2 + \dot Y_{\rm as}^2 + \dot Z_{\rm as}^2).
\eea
This completes the specification 
of the new family of solutions of fireball
hydrodynamics with ellipsoidal symmetry.
The form of the dimensionless scaling function 
${\cal T}(s)$ can be chosen freely from among the positive
functions of a non-negative variable. 
This freedom corresponds  
to a freedom in the specification of the 
initial conditions.
Thus (uncountably) infinite new solutions of 
NR hydrodynamics are found. 

{\it Self-similarity of the elliptic hydro of solutions} --- 
Each of these new hydrodynamical solutions  is scale invariant:
\bea
	{\bf r}^\prime & = & 
		(r_x \frac{X_0}{X}, r_y \frac{Y_0}{Y}, r_z \frac{Z_0}{Z}), \\
	n\left(t, {\bf r}\right) &  = & 
		  n(t_0, {\bf r}^\prime)\, 
		  \left(\frac{ X_0 Y_0 Z_0}{ X Y Z}\right), 
		  \label{e:nscale}  \\
	v_x\left(t,{\bf r}\right) 
		  &  = &  v_x (t_0, {\bf r}^\prime)\frac{\dot X}{\dot X_0},  ...
			\label{e:vscale} \\ 
	T\left(t, {\bf r}\right)
		  &  = & T(t_0, \bvec r^\prime)
		  \left( \frac{ X_0 Y_0 Z_0}{ X Y Z}
		  \right)^{1/\kappa}. \label{e:tscale}
\eea
Scale invariance of these solutions is equivalent to their self-similarity.
The profile functions depend on time only
through the scale parameters $(X,Y,Z)$ and 
on the coordinates only through the scale parameter $s$.

{\it Limiting cases} ---
The various physically interesting limiting cases fall into two classes.
The first class of limiting cases corresponds to various additional symmetry
properties imposed on the scaling variable of eq.~(\ref{e:scale}).
The second class of limiting cases corresponds to
various choices of ${\cal T}(s)$, the scaling function of the temperature. 
The Lagrangian equations  of motion for
the scale parameters, eqs. ~(\ref{e:lag}) as well as the shape of the
flow velocity field, eq.~(\ref{e:v}), are the same for
all the choices of ${\cal T}(s)$.  
Trivial prefactors
are given in eqs.~(\ref{e:n},\ref{e:T}).
Hence these equations will not be
repeated in the forthcoming discussion. We provide some physically interesting
examples for  
the matching pair of density and temperature  scaling functions 
$\left(\nu,{\cal T}\right)$ 
that satisfy eq.~(\ref{e:nutau}).

{\it  Spherically symmetric family of solutions} --- 
By assuming that initially all the scale parameters as well as 
their time derivatives are 
equal, $X_0 = Y_0 = Z_0 \equiv R_0$ 
and $\dot X_0 = \dot Y_0 = \dot Z_0 \equiv \dot R_0$,
the ellipsoidal family of solutions reduces 
to the spherical family of solutions  of ref.~\cite{cssol}
with a scale 
parameter $ X = Y = Z \equiv R$.  
The scaling variable and the equations of motion simplify to
\be
	s = \frac{\bf r^2}{R^2}, \quad
 	R\ddot R = \frac{T_i}{m} \left(\frac{R_0^3}{R^3}\right)^{1/\kappa}.
\ee
This family 
generalizes the Zim\'anyi-Bondorf-Garpman (ZGB) solution~\cite{jnr},
the spherical Gaussian solution~\cite{nr}, 
and the Buda-Lund type of hydro solutions~\cite{cssol}
to  arbitrary scaling functions ${\cal T}(s) > 0$ and
equation of state parameters $\kappa > 0$.
If $\dot R_0 = 0$, 
and the asymptotic velocity of the expansion  is
$\dot R_{\rm as}^2 = \langle u \rangle^2 = T_i/m$, 
and if $\kappa = 3/2$, the equation of motion of 
the scale parameter is solved in a simple form~\cite{cssol}  as 
$R^2 = R_0^2 + \langle u \rangle^2 (t - t_0)^2$. 

{\it Cylindrically symmetric family of solutions} ---
By imposing cylindrically symmetric initial conditions, 
$X_0 = Y_0 = R_{t0}$, and
$\dot X_0 = \dot Y_0 = \dot R_{t0}$, 
one finds that $X = Y = R_t$, 
as the equations of motion preserve
cylindrical symmetry.  Introducing $r_t = \sqrt{r_x^2 + r_y^2}$, 
the scale parameter $s$ and the equations
of motion for the longitudinal and the transverse scales read as
\bea
	s & = &\frac{r_t^2}{R_t^2} + \frac{r_z^2}{Z^2}, 
	\quad 
	R_t \ddot R_t \, = \, Z \ddot Z \, =\, \frac{T_i}{m}
		\left(\frac{R_{t0}^2 Z_0}{R_t^2 Z_0}\right)^{1/\kappa}.
\eea
These generalize the equations of motion for scales of the
cylindrically symmetric, De - Garpman - Sperber - Bondorf
- Zim\'anyi (DGSBZ) solution of ref.~\cite{jde} to arbitrary 
$0 < \kappa \ne 3/2$ and to arbitrary ${\cal T}(s) >0$.

{\it One dimensional expansions} --- 
The equations of motion of parameters $(X,Y,Z)$ has been studied
for the case of $\kappa = 3/2 $ in ref.~\cite{ellsol}. 
Although the expansion is generally 3 dimensional, 
a big initial compression in one of
the directions ($r_z$) was shown to result in an effectively 1 dimensional 
expansion in this direction, corresponding to Landau type initial conditions.  
In this case, an analytic solution for the variable $Z$ is 
given in eqs. (23-25) of ref.~\cite{ellsol},
and the conditions of validity of this 
approximation were given there by eqs. (26-28).
These equations can be further simplified during the late stage of
the expansion, when acceleration effects are small. In this limiting case,
both eqs.~(\ref{e:lag}) and the conservation of
energy are satisfied by the asymptotic solution:
\bea
	\dot X_a                 & \simeq & \dot X_0, \quad 
	\dot Y_a                 \, \simeq \, \dot Y_0, \quad
	\frac{1 }{2} m \dot Z_a^2\, \simeq \, \frac{3}{2} T_i, \\
	X(t)			& \simeq & X_0 + \dot X_a t,\label{e:aX}\\ 
	Y(t)			& \simeq & Y_0 + \dot Y_a t,\label{e:aY}\\
	Z(t)			& \simeq & Z_0 + \dot Z_a t.\label{e:aZ}
\eea
	For simplicity, here we utilized $t_0 = 0$ .
	This asymptotic approximate  solution is valid if the conditions of
	validity of the 1 dimensional expansion given in ref.~\cite{ellsol}
	are satisfied simultaneously with the following
	constraints: $Z_0 \ll \dot Z_a t$, $ X_0 \gg \dot X_0 t$
	and $ Y_0 \gg \dot Y_0 t$.
	Alternatively $ \sqrt{m/(3 T_i)} Z_0 \ll t \ll 
	\min( X_0/\dot X_0, Y_0/\dot Y_0) $.
%	If the expansion in the beam direction is large as compared to the 
%	initial $Z_0$ and the expansion in the transverse directions is 
%	small, as compared to the initial transverse coordinate values,	
	Under these conditions
%	the transverse and the longitudinal dynamics decouple and 
	all
	the initial internal energy is converted into kinetic energy
	in the longitudinal direction, while the kinetic energy
	in the transverse components is conserved during the time
	evolution. For $\dot X_0 = \dot Y_0 = 0$, a one dimensional expansion 
	is obtained.

Various choices for the shape of the temperature scaling function 
${\cal T}(s)$
generate interesting forms of the hydrodynamical solutions in all the
ellipsoidal, cylindrical, spherical  or 1 dimensionally expanding classes. 

{\it  Gaussian solutions} ---
The simplest possible choice for the temperature scaling function 
is  ${\cal T}(s) = 1$. 
After a trivial scale transformation, $(X,Y,Z) \rightarrow 
\sqrt{\frac{T_i}{T_0} } (X,Y,Z)$, 
\begin{eqnarray}
	\nu(s) & = &  
    		\mathrm{exp}\left( - s/2 \right),  \quad {\cal T}(s) = 1.
		\label{e:ngauss} 
\end{eqnarray}
The density profiles are Gaussians and 
the temperature distribution becomes spatially homogeneous, 
as follows from eq.~(\ref{e:scale}), and 
we recover the elliptic Gaussian solutions 
described in refs.~\cite{ellsol,csell}.
If freeze-out happens at a constant value of the local temperature,
$T(t,{\rm {\bf r}}) = T_f$, these 
Gaussian hydrodynamical solutions corresponds
to a sudden freeze-out at a constant time, $t = t_f$, in the whole volume 
of the fireball. 
Remarkable features of this model are that {\it i)}
the slope parameters of the transverse momentum spectra increase 
linearly with mass, with the coefficient of linearity depending
on the relative direction to that of the 
 impact parameter, {\it ii)} the parameters
of the two-particle Bose-Einstein correlation functions oscillate as
a function of the angle between the event plane and the transverse momentum
of the pair~\cite{csell}.

{\it  De - Garpman - Sperber - Bondorf - Zim\'anyi solution} ---
One can require that the temperature and the density profiles 
are described as different powers of the same profile function.
Such a similarity is achieved if the temperature 
and the corresponding density profiles are 
\bea
	{\cal T}(s) & = & \left( 1 - s \right)  \Theta(1 -s), \\
	{\nu}(s) & = & \left( 1 - s \right)^\alpha \Theta(1 -s), \quad
	\alpha = T_i/(2T_0) - 1 .
\eea
These profile functions generalize the spherical 
ZGB solution~\cite{jnr,cssol}, 
and the cylindrically symmetric 
DGSBZ solution~\cite{jde}
to asymmetric ellipsoids and  arbitrary values of 
$\kappa$. 
The physical meaning of the parameter
$\alpha$ is  
determined here, in terms of 
the total  initial internal energy $T_i$  and 
the initial central value $T_0$ of the temperature field.
\begin{figure}[tbp]
\vspace*{7.5cm}
%\vspace*{5.0cm}
\includegraphics{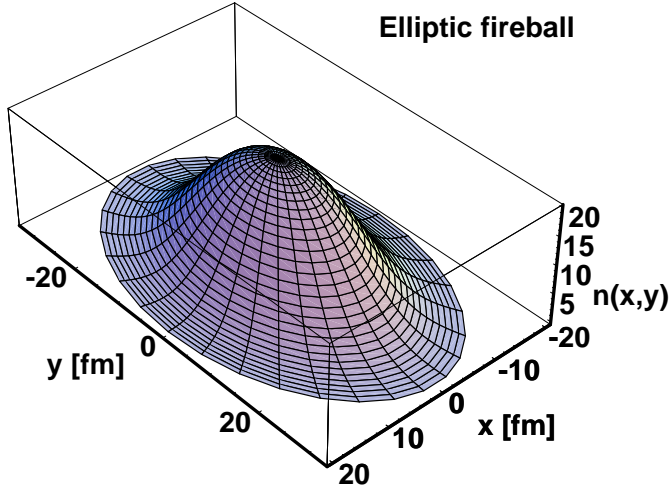}
\vspace{-1.0cm}
\caption{\label{f:gauss} An approximately Gaussian 
BL-H fireball profile is shown 
for $X = 5$ fm, $Y= 8$ fm, $m / T_0 = 1$, $\langle u \rangle = 0.7$, 
$\langle \Delta T/T \rangle = 0.1$. The vertical scale is arbitrary, 
the longitudinal coordinate is $r_z = 0$. For the time dependence see eqs.~(\ref{e:lag}).
}
\vspace{-0.5cm}
\end{figure}

\begin{figure}[tbp]
\vspace*{7.0cm}
\includegraphics{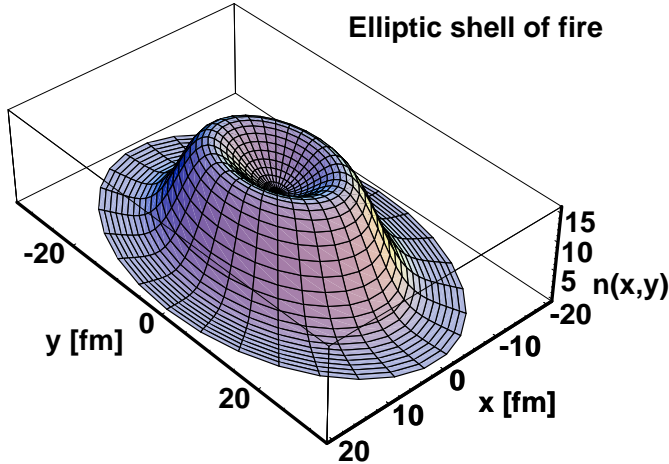}
\vspace{-1.0cm}
\caption{\label{f:shell}
A BL-H shell profile with $X = 5$ fm,
$Y= 8$ fm, $m / T_0 = 1$, $\langle u \rangle = 0.5$, 
$\langle \Delta T/T \rangle = 0.71$, otherwise as Fig. 1.
}
\end{figure}

{\it  Elliptic Buda - Lund solutions ---}
The Buda - Lund hydro model (BL-H) was developed for the description of
single-particle spectra and two-particle Bose-Einstein correlation functions
in high energy heavy ion collisions at CERN SPS~\cite{3d,qm95}. The BL-H
attempted to characterize the temperature, density and flow
fields by their means and variances only, its essential property is
that it simultaneously involves a temperature gradient parameter and
a Hubble-like flow profile. In its original form, the model was
cylindrically symmetric and the (longitudinal) flow profile was 
relativistic~\cite{3d,qm95}.  
BL-H type of exact hydro solutions correspond to  
the scaling functions 
\bea
	{\cal T}(s)  & = & \frac{1}{1 + b s}, 
		\quad b \,  = \, 
		\frac{1}{2}\langle \frac{\Delta T}{T}\rangle, \\
 	{\nu}(s)   & = & \left( 1 + b s \right)
	\exp\left[ - \frac{T_i}{2 T_0} ( s   + b s^2/2)  \right].
\eea
The dimensionless parameter $b$ is interpreted as a measure of the
transverse temperature inhomogeneity~\cite{3d,qm99}. 
Regardless of the symmetry classes, 
the BL-H density profile has an approximately
Gaussian, conventional shape if $b < T_i/( 2 T_0)$.
On the other hand, the density
profile looks like an ellipsoidal ring of fire, with a density minimum
at the center and a density pile-up on the surface, if 
$b > T_i/T_0$.
In the spherically symmetric case, similar morphological classes of   
BL-H solutions were found in ref~\cite{cssol}. 
Introducing the notation $T_i = m \langle u\rangle^2$, one finds that 
$\frac{m \langle u \rangle^2}{T_0} > \langle \frac{\Delta T}{T}\rangle$
yields ellipsoidal, expanding fireball BL-H solutions, Fig.~\ref{f:gauss}. 
A detailed analysis of 
correlations and spectra in Pb + Pb collisions at 
CERN SPS indicated~\cite{qm99}
such a behavior, corresponding to big and expanding fireballs.
On the other hand, the analysis of correlations and spectra of
 $h + p$ reactions at CERN SPS  indicated ~\cite{csrev,krev,na22}
small transverse flow and big transverse temperature inhomogeneity,
$\frac{m \langle u \rangle^2}{T_0} <
\langle \frac{\Delta T}{T}\rangle$. This case corresponds to 
the formation of a shell of 
fire. Such an example is shown in Fig.~\ref{f:shell}. 

{\it  Fireballs with density waves ---}
Here we show that a simple choice of the temperature scaling 
function can lead to periodically oscillating temperature and density 
waves in exact solutions of NR hydrodynamics.
A pair of oscillating scaling functions is, for example 
\bea
	{\cal T}(s) & =  &({1 + \alpha \cos\beta s})^{-1}, \\
	\nu(s) & = & 
		\left(1 + \alpha \cos\beta s\right)
		\exp\left[ - \frac{T_i}{2T_0} (s - 
		\frac{\alpha}{\beta} \sin\beta s) \right]. \nonumber \\
	&&
\eea
where parameters $(\alpha,\beta)$ correspond to  
the amplitude and the period of the oscillations, respectively, as 
shown in Fig. 3.  As time evolves, the
shapes of the ellipsoids change in coordinate space: 
the bigger the initial compression, 
the faster the expansion in that direction, corresponding to
eqs.~(\ref{e:lag}). 

\begin{figure}[tbp]
\null
\vspace*{7.0cm}
\includegraphics{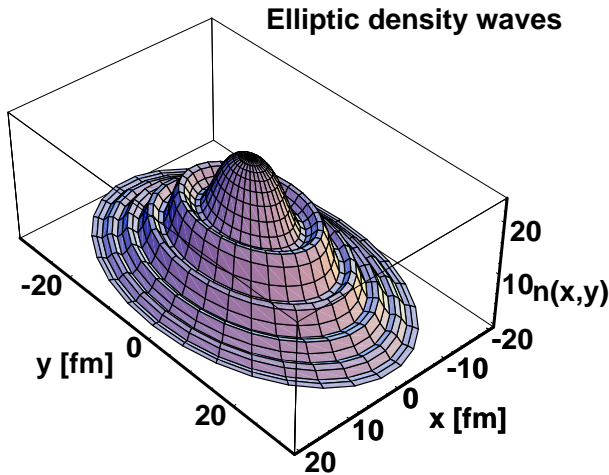}
\vspace{-1.0cm}
\caption{An oscillating wave-like density profile is obtained, for $X = 5$ fm,
$Y= 8$ fm, $T_i/ T_0 = 1$, 
$\alpha= 0.2$, $\beta =2.0$.
}
\end{figure}
{\it Connection with other exact solutions of fireball hydrodynamics} --
Before summarizing the results let us also relate the presented family
of exact solutions of hydrodynamics to presently known, exact
solutions of relativistic hydrodynamics. 
One of the most well known relativistic hydrodynamical solution
is the 1+1 dimensional Landau-Khalatnikov solution, described in
refs.~\cite{Landau:1953gs}, \cite{Khalatnikov} and \cite{Belenkij:1956cd}.
The initial condition here is a uniformly heated piece of matter initially at rest,
the equation of state is that of an ultrarelativistic gas with three degrees of 
freedom, but expanding only 1 + 1 dimensions. Landau has shown that such expansions
lead to an approximately Gaussian rapidity distribution. This Landau-Khalatnikov solution
is presently the only known exact solution of relativistic hydrodynamics that 
describes an exploding fireball with relativistic acceleration. These solutions
are, however, not self-similar, and not explicit, one dimensional  solutions, hence they are not
easily related to the exact, explicit, selfsimilar and accelerating non-relativistic
solutions described in the present manuscript. 

However, it is interesting to note, that in the late time limit, the
acceleration of the scale parameters vanishes in the non-relativisitic self-similar
solutions described here, as given by eqs. (\ref{e:aX}-\ref{e:aZ}).
Hence for very late time, the velocity field becomes spherically symmetric,
${\bf v} \rightarrow {\bf r}/t$, 
which corresponds, within the lightcone, to a well known solution of relativistic hydrodynamics: 
$u^\mu(x) = x^\mu/\tau$, which is in the 1+1 dimensional case the reknowned velocity distribution
of the Hwa-Bjorken solution  of relativistic
hydrodynamics, given in refs.~\cite{Hwa:1974gn},\cite{Bjorken:1982qr},
while in 1 + 3 dimensions ${\bf v}  = {\bf r}/t$ or $u^\mu(x) = x^\mu/\tau$ 
is referred to as the Hubble flow, 
as it corresponds to the flow velocity field of galaxies in an expanding Friedmann
universe.
These solutions are boost-invariant and both in the  1+1 dimensional Hwa-Bjorken
and in the 1+3 dimensional Hubble case, corresponding to a flat rapidity distribution. 
The initial boundary conditions are given only within a lightcone, on a $\tau = \tau_0$
boost-invariant hypersurface, the equation of state is characterized by a
constant or piecewise constant speed of sound, and the solutions are explicit, accelerationless
and self-similar flows.

In ref.~\cite{Csorgo:2003rt} these boost-invariant self-similar solutions were generalized
to 1+1 dimensional, non-boost invariant, self-similar, accelerationless solutions,
using a broad class of equations of state: $\epsilon = m n + \kappa p$, $p = nT$.
The initial conditions were given, similarty to the case of the Hwa-Bjorken
solution, on a boost-invariant hypersurface with  $\tau = \tau_0$,
but assuming an inhomogeneous initial temperature profile and a corresponding
matching initial density profile.
In the same paper this solution is extended to axially symmetic, three dimensionally
expanding fireballs, that represent  the late stages of central 
heavy ion collisions. However, for non-central collisions, axial symmetry is too
restricitve and due to this reason we have generalized these solutions 
for ellipsoidally expanding relativistic fireballs, and 
the 1+1 dimensional solution, the 1+3 dimensional axially symmetric solution
and the 1+3 dimensional ellipsoidal solutions were written up also in a series of papers
for the 3rd Budapest Winter School on Heavy Ion Physics, published in 
refs. \cite{Csorgo:2002ki},\cite{Csorgo:2002bi},\cite{Csorgo:2003ry}.
These self-similar, relativistic hydrodynamical solutions, in particular,
the ellipsoidal expansion described in ref.~\cite{Csorgo:2003ry},
correspond exactly to the late time limit of the non-relativistic solutions
described here. See ref.~\cite{Csorgo:2003ry} for further details on such
a correspondence as well as for a more complete
list of citations on early papers on the Hwa-Bjorken solution.

It is worthwhile to mention that there are additional, recently found 
exact solutions of relativistic hydrodynamics that
find accelerationless generalized Hubble type of solutions, with direction
dependent Hubble constants. The first of these class of solutions has been
found by B\'{\i}r\'o in refs. \cite{Biro:1999eh},\cite{Biro:2000nj},
for cylindrically symmetric expansions at the softest point of the equation of state.
Sinyukov and Karpenko recently published a solution which can be considered
as the generalization of refs. \cite{Biro:1999eh},\cite{Biro:2000nj},
as well as that of ref.  \cite{Csorgo:2003ry}, for the pre-asymptotic,
accelerationless stage of the expansion, when the scales expand
already linearly in time, corresponding to eqs. (\ref{e:aX}-\ref{e:aZ}),
but before the time period when the off-sets $(X_a , Y_a, Z_a)$
can be considered negligibly small.
The B\'{\i}r\'o as well as the Sinyukov-Karpenko solutions 
are also self-similar and accelerationless relativistic solutions,
their shortcoming is that they are valid in the medium only when
the pressure is a temperature independent constant, or,
when the boundary condition is an expansion to the vacuum,
they are obtained only for the special "dust" equation of state, that
corresponds to  a uniformly vanishing pressure, $p=p_0= 0$.
These solutions can thus be considered as a pre-asymptotic,
relativistic solutions corresponding to very late stages of the fireball
expansions. 

This brief review of known exact solutions of fireball hydrodynamics in
the relativistic kinematic domain indicates, that we find a correspondence
between the late stages of the presented exact, accelerating, self-similar
solutions of  non-relativistic, ellipsoidally symmetric fireballs and the
known class of relativistic but accelerationless, self-similar, ellipsoidally symmetric 
solutions of hydrodynamics. The missing link between them is
the class of relativistic, accelerating, explicit
solutions of hydrodynamics that use a realistic equation of state. 
The search for this missing class of exact and explicit solutions has 
been started and the first results will soon be reported elsewhere.

{\it Summary ---} The non - relativistic hydrodynamical problem
has been solved for expanding fireballs with ellipsoidal symmetry
for the class of self-similar expansions. 
The flow velocity distribution is a generalized Hubble field in all the cases.
An exact solution is assigned 
to each positive, integrable function of a non-negative variable. 
The time evolution of the $(X,  Y, Z)$ scale parameters
corresponds to a Hamiltonian motion of a mass point 
in a non-central, repulsive potential.
The density profiles may be
of fireball type, or they may form one or more shells of fire.
For initial conditions with higher symmetry, one dimensional,
cylindrical and spherical expansions are obtained.

These results provide analytic insight into the time evolution of expanding
fireballs with nontrivial, ellipsoidally symmetric morphology, a 
non-polynomially hard algorithmic problem 
that is difficult to solve even numerically on von Neumann type computers.

{\it Acknowledgments:} Thanks are due to  L. P. Csernai,
Y. Hama, F. Grassi, M. Gyulassy, T. Kodama, B. Luk\'acs and J. Zim\'anyi
for inspiring discussions. This work was supported by 
OTKA grants T026435, T034296, T 038406, NWO - OTKA  grant N 25487, by  an 
NSF - MTA - OTKA grant INT0089462, the US DOE grant 
DE - FG02 - 93ER40764 and by the NATO PST.CLG.980086 grant.

\medskip
\end{document}